%% file: main.tex
\documentclass[runningheads]{llncs}
\usepackage[T1]{fontenc}
\usepackage{graphicx}
\usepackage{multirow}
\usepackage{booktabs}
\usepackage{gensymb}
\usepackage{color,xcolor}
\usepackage{xspace}
\usepackage{epstopdf}
\usepackage{algorithm}
\usepackage{algorithmic}
\usepackage{amsmath,amsfonts}
\usepackage{amssymb,amsopn}
\usepackage{bm}
\usepackage{url}
\usepackage{subfigure}
\usepackage[misc]{ifsym}  
\usepackage[colorlinks=true,linkcolor=blue,citecolor=blue,urlcolor=blue]{hyperref}

\newcommand{\tabincell}[2]{\begin{tabular}{@{}#1@{}}#2\end{tabular}}
\newcommand{\mat}[1]{\boldsymbol{#1}} 
\newcommand{\freeseed}{{FreeSeed}\xspace}
\newcommand{\dudofreeseed}{FreeSeed\textsuperscript{\textsc{dudo}}\xspace}
\newcommand{\freenet}{{FreeNet}\xspace}
\newcommand{\seednet}{{SeedNet}\xspace}
\newlength\savewidth\newcommand\shline{
\noalign{\global\savewidth\arrayrulewidth
\global\arrayrulewidth1pt}\hline\noalign{\global\arrayrulewidth\savewidth}}

\begin{document}
\title{\freeseed: Frequency-band-aware and Self-guided Network for Sparse-view CT Reconstruction}
\titlerunning{Frequency-band-aware and Self-guided Network for Sparse-view CT}

\author{Chenglong Ma\inst{1} \and
Zilong Li\inst{1} \and 
Junping Zhang\inst{1} \and 
Yi Zhang\inst{2}  \and \\
Hongming Shan\inst{1}}
\authorrunning{C. Ma et al.}

\institute{Fudan University \\
\email{clma22@m.fudan.edu.cn, hmshan@fudan.edu.cn} 
\and
Sichuan University 
}

\maketitle
\input{secs/abstract}
\input{secs/introduction}

\input{secs/method}

\input{secs/experiments}
\input{secs/conclusion}


\input{secs/bbl}
\newpage
\section*{Supplementary Materials}
\renewcommand*{\thefigure}{S\arabic{figure}}
\setcounter{figure}{0}
\input{secs/supp}

\end{document}

%% file: secs/abstract.tex
\begin{abstract}
Sparse-view computed tomography (CT) is a promising solution for expediting the scanning process and mitigating radiation exposure to patients, the reconstructed images, however, contain severe streak artifacts, compromising subsequent screening and diagnosis. Recently, deep learning-based image post-processing methods along with their dual-domain counterparts  have shown promising results. However, existing methods usually produce over-smoothed images with loss of  details due to (\textbf{1}) the difficulty in accurately modeling the artifact patterns  in the image domain, and (\textbf{2}) the equal treatment of each pixel in the loss function.
To address these issues, we concentrate on the image post-processing and propose a simple yet effective FREquency-band-awarE and SElf-guidED network, termed \freeseed, which can effectively remove artifact and recover missing detail from the contaminated sparse-view CT images.
Specifically, we first propose a frequency-band-aware artifact modeling network (\freenet), which learns artifact-related frequency-band attention in Fourier domain for better modeling the globally distributed streak artifact on the sparse-view CT images. 
We then introduce a self-guided artifact refinement network (\seednet), which leverages the predicted artifact to assist \freenet in continuing to refine the severely corrupted details. 
Extensive experiments demonstrate the superior performance of \freeseed and its dual-domain counterpart over the state-of-the-art sparse-view CT reconstruction methods. Source code is made available at \url{https://github.com/Masaaki-75/freeseed}.

\end{abstract}

%% file: secs/introduction.tex
\section{Introduction}
X-ray computed tomography (CT) is an established diagnostic tool in clinical practice; however, there is growing concern regarding the increased risk of cancer induction associated with X-ray radiation exposure~\cite{ct-outlook}.
Lowering the dose of CT scans has been widely adopted in clinical practice to address this issue, following the “as low as reasonably achievable” (ALARA) principle in medical community~\cite{alara}. Sparse-view CT is one of the effective solutions, which reduces the radiation by only sampling part of the projection data for image reconstruction. Nevertheless, images reconstructed by the conventional filtered back-projection (FBP) present severe artifacts, thereby compromising its clinical value.

In recent years, the success of deep learning has attracted much attention in the field of sparse-view CT reconstruction. Existing learning-based approaches mainly include image-domain methods~\cite{residual-manifold,fbpconvnet,ddnet} and dual-domain methods~\cite{drone,dudonet,dudotrans}, both involving image post-processing to restore a clean CT image from the low-quality one with streak artifacts. For the image post-processing, residual learning~\cite{resnet} is often employed to encourage learning the artifacts hidden in the residues, which has become a proven paradigm for enhancing the performance~\cite{residual-manifold,fbpconvnet,sinogram-interp,drone}. 
Unfortunately, existing image post-processing methods may fail to model the globally distributed artifacts within the image domain. They can also produce over-smoothed images due to the lack of differentiated supervision for each pixel. In this paper, we advance the image post-processing to benefit both classical image-domain methods and the dominant dual-domain ones.

\begin{figure}[t]
\centering
\subfigure[$N_\mathrm{v}=18$]
{\includegraphics[width=0.24\linewidth]{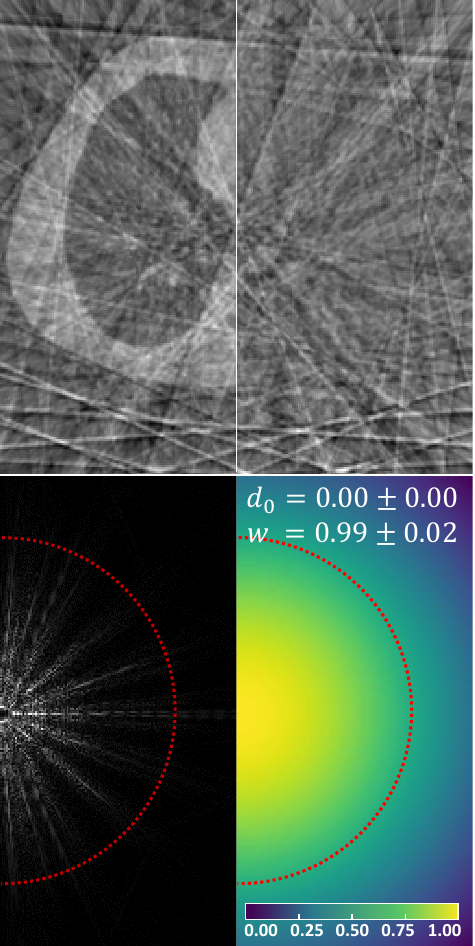}}
\label{fig:amp-art-18}
\subfigure[$N_\mathrm{v}=36$]
{\includegraphics[width=0.24\linewidth]{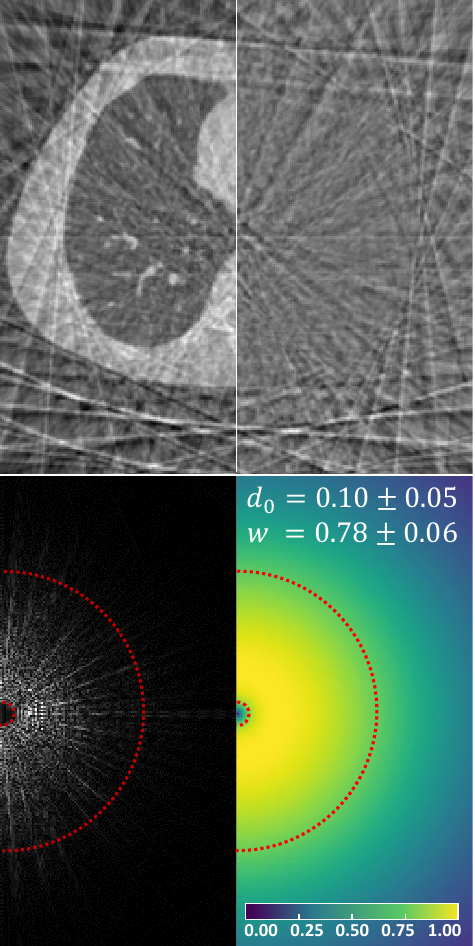}}
\label{fig:amp-art-36}
\subfigure[$N_\mathrm{v}=72$]
{\includegraphics[width=0.24\linewidth]{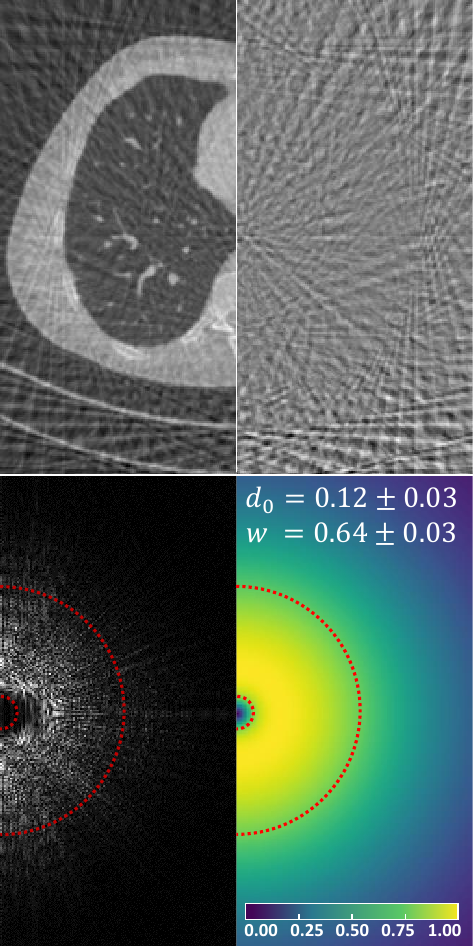}}
\label{fig:amp-art-72}
\subfigure[$N_\mathrm{v}=144$]
{\includegraphics[width=0.24\linewidth]{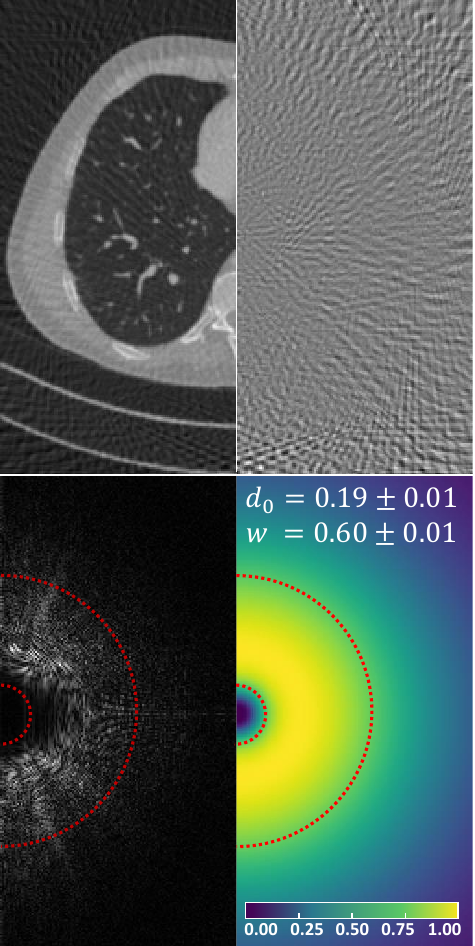}}
\label{fig:amp-art-144}
\caption{First row: sparse-view CT images (left half) and the corresponding artifacts (right half); Second row: real Fourier amplitude maps of artifacts (left half) and the learned band-pass attention maps (right half, with inner radius and bandwidth respectively denoted by $d_0$ and $w$. Values greater than 0.75 are bounded by red dotted line) given different number of views $N_\mathrm{v}$.}
\label{fig:amp-artifact}
\end{figure}

\noindent\textbf{Motivation.}\quad 
We view the sparse-view CT image reconstruction as a two-step task: artifact removal and detail recovery. For the former, few work has investigated the fact that the artifacts exhibit similar pattern across different sparse-view scenarios, which is evident in \emph{Fourier domain} as shown in Fig.~\ref{fig:amp-artifact}: they are aggregated mainly in the mid-frequency band and gradually migrate from low to high frequencies as the number of views increases. 
Inspired by this, we propose a frequency-band-aware artifact modeling network (\freenet) that learns the artifact-concentrated frequency components to remove the artifacts efficiently using learnable band-pass attention maps in Fourier domain.

While Fourier domain band-pass maps help capture the pattern of the artifacts, restoring the image detail contaminated by strong artifact may still be difficult due to the entanglement of artifacts and details in the residues. We therefore propose a self-guided artifact refinement network (\seednet) that provides supervision signals to aid \freenet in refining the image details contaminated by the artifacts. With these novel designs, we introduce a simple yet effective model termed FREquency-band-awarE and SElf-guidED network (\freeseed), which enhances the reconstruction by modeling the pattern of artifacts from a frequency perspective and utilizing the artifact to restore the details. \freeseed achieves promising results with only image data and can be further enhanced once the sinogram is available. 

Our contributions can be summarized as follows: (\textbf{1}) a novel frequency-band-aware network is introduced to efficiently capture the pattern of global artifacts in Fourier domain among different sparse-view scenarios; (\textbf{2}) to promote the restoration of heavily corrupted image detail, we propose a self-guided artifact refinement network that ensures targeted refinement of the reconstructed image and consistently improves the model performance across different scenarios; and (\textbf{3}) quantitative and qualitative results demonstrate the superiority of \freeseed over the state-of-the-art sparse-view CT reconstruction methods.

%% file: secs/method.tex
\section{Methodology}
\begin{figure}[t]
\centering
\includegraphics[width=\linewidth]{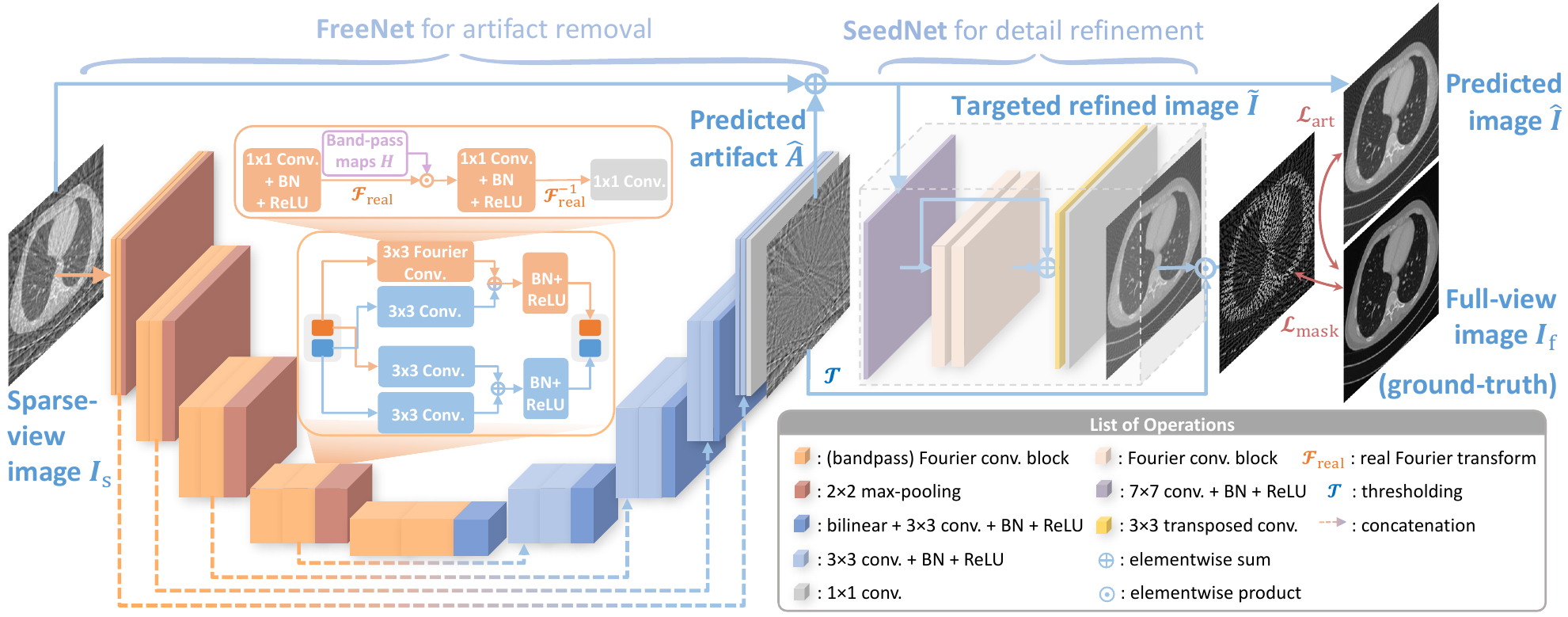}
\caption{Overview of the proposed \freeseed.}
\label{fig:model}
\end{figure}

\subsection{Overview}
Given sparse-view sinogram with projection views $N_\mathrm{v}$, let $\mat{I}_\mathrm{s}$ and $\mat{I}_\mathrm{f}$ denote the directly reconstructed sparse- and full- view images by FBP, respectively. In this paper, we aim to construct an image-domain model to effectively recover $\mat{I}_\mathrm{s}$ with a level of quality close to $\mat{I}_\mathrm{f}$.

The proposed framework of \freeseed is depicted in Fig.~\ref{fig:model}, which mainly consists of two designs: \freenet that learns to remove the artifact and is built with band-pass Fourier convolution block that better captures the pattern of the artifact in Fourier domain; and \seednet as a proxy module that enables \freenet to refine the image detail under the guidance of the predicted artifact.
Note that \seednet is involved only in the training phase, additional computational cost will not be introduced in the application. The parameters of \freenet and \seednet in \freeseed are updated in an iterative fashion. 

\subsection{Frequency-band-aware Artifact Modeling Network}
To learn the globally dispersed artifact, \freenet uses band-pass Fourier convolution block as the basic unit to encode artifact from both spatial and frequency aspect. Technically, Fourier domain knowledge is introduced by fast Fourier convolution (FFC)~\cite{ffc}, which benefits from the non-local receptive field and has shown promising results in various computer vision tasks~\cite{lama,swinfir}. The features fed into FFC are split evenly along the channel into a spatial branch composed of vanilla convolutions and a spectral branch that applies convolution after real Fourier transform, as shown in Fig.~\ref{fig:model}.

Despite the effectiveness, simple Fourier unit in FFC could still preserve some low-frequency information that may interfere with the learning of artifacts, which could fail to accurately capture the banded pattern of the features of sparse-view artifacts in frequency domain. To this end, we propose to incorporate learnable band-pass attention maps into FFC. Given an input spatial-domain feature map $\mat{X}_\mathrm{in}\in\mathbb{R}^{C_\mathrm{in}\times H\times W}$, the output $\mat{X}_\mathrm{out}\in\mathbb{R}^{C_\mathrm{out}\times H\times W}$ through the Fourier unit with learnable band-pass attention map is obtained as follows:
\begin{align}
    \mat{Z}_\mathrm{in} = \mathcal{F}_\mathrm{real}\{\mat{X}_\mathrm{in}\}, \quad\quad
    \mat{Z}_\mathrm{out} = 
    f\left(
    \mat{Z}_\mathrm{in}\odot\mat{H}
    \right), \quad\quad
    \mat{X}_\mathrm{out} = \mathcal{F}^{-1}_\mathrm{real}\{\mat{Z}_\mathrm{out}\},
\end{align}
where $\mathcal{F}_\mathrm{real}$ and $\mathcal{F}^{-1}_\mathrm{real}$ denote the real Fourier transform and its inverse version, respectively. $f$ denotes vanilla convolution. “$\odot$” is the Hadamard product. Specifically, for $c$-th channel frequency domain feature $\mat{Z}^{(c)}_\mathrm{in}\in\mathbb{C}^{U\times V} (c=1,...,C_\mathrm{in})$, the corresponding band-pass attention map $\mat{H}^{(c)}\in\mathbb{R}^{U\times V}$ is defined by the following Gaussian transfer function:

\begin{align}
    \mat{H}^{(c)} &= \exp\left[-\left(
    {\dfrac{\mat{D}^{(c)2}-d_0^{(c)2}}
    {w^{(c)}\mat{D}^{(c)}+\epsilon}
    }\right)^2\right],
    \label{eq:band-pass-gaussian}\\
    \mat{D}^{(c)}_{u,v} &= 
    \sqrt{\dfrac{(u-U/2)^2+(v-V/2)^2}{\max_{u',v'}(u'-U/2)^2+(v'-V/2)^2}},\label{eq:distance-map}
\end{align}
where $\mat{D}^{(c)}$ is the $c$-th channel of the normalized distance map with entries denoting the  distance from any point $(u,v)$ to the origin. $w^{(c)}>0$ and $d^{(c)}_0\in[0,1]$ are two learnable parameters representing the bandwidth and the normalized inner radius of the band-pass map and are initialized as 1 and 0, respectively. $\epsilon$ is set to $1\times10^{-12}$ to avoid division by zero. The right half part of the second row of Fig. \ref{fig:amp-artifact} shows some samples of the band-pass maps.

The pixel-wise difference between the predicted artifact $\widehat{\mat{A}}$ of \freenet and the groundtruth artifact $\mat{A}_\mathrm{f}=\mat{I}_\mathrm{f}-\mat{I}_\mathrm{s}$ is measured by $\ell_2$ loss:
\begin{align}
    \mathcal{L}_\mathrm{art} = \Vert\mat{A}_\mathrm{f}-\widehat{\mat{A}}\Vert_2.
    \label{eq:loss-art}
\end{align}

\subsection{Self-guided Artifact Refinement Network}
Areas heavily obscured by the artifact should be given more attention, which is hard to achieve using only \freenet. Therefore, we propose a proxy network \seednet that provides supervision signals to focus \freenet on refining the clinical detail contaminated by the artifact under the guidance of the artifact itself. \seednet consists of residual Fourier convolution blocks. 
Concretely, given sparse-view CT images $\mat{I}_\mathrm{s}$, \freenet predicts the artifact $\widehat{\mat{A}}$ and restored image $\widehat{\mat{I}}=\mat{I}_\mathrm{s}-\widehat{\mat{A}}$; the latter is fed into \seednet to produce targeted refined result $\widetilde{\mat{I}}$.
To guide the network on refining the image detail obscured by heavy artifact, we design the transformation $\mathcal{T}$ that turns $\widehat{\mat{A}}$ into a mask $\mat{M}$ using its mean value as threshold: $\mat{M} = \mathcal{T}(\widehat{\mat{A}})$, and define the following masked loss for \seednet: 
\begin{align}
    \mathcal{L}_\mathrm{mask} = \Vert(\mat{I}_\mathrm{f}-\widetilde{\mat{I}})\odot\mat{M}\Vert_2.
    \label{eq:loss-mask}
\end{align}

\subsection{Loss Function for \freeseed}
\freenet and \seednet in our proposed \freeseed are trained in an iterative fashion, where \seednet is updated using $\mathcal{L}_\mathrm{mask}$ defined in Eq. \eqref{eq:loss-mask}, and \freenet is trained under the guidance of the loss defined in Eq. \eqref{eq:loss-model}, with  $\alpha>0$ set as 1 for the best performance. The pseudo code for training process and the exploration on the selection of $\alpha$ can be found in our Supplementary Material. Once the training is complete, \seednet can be dropped and the prediction is done by \freenet.
\begin{align}
    \mathcal{L}_\mathrm{total} = \mathcal{L}_\mathrm{art} + \alpha\mathcal{L}_\mathrm{mask}.
    \label{eq:loss-model}
\end{align}

\begin{figure}[htbp]
\centering
\includegraphics[width=\linewidth]{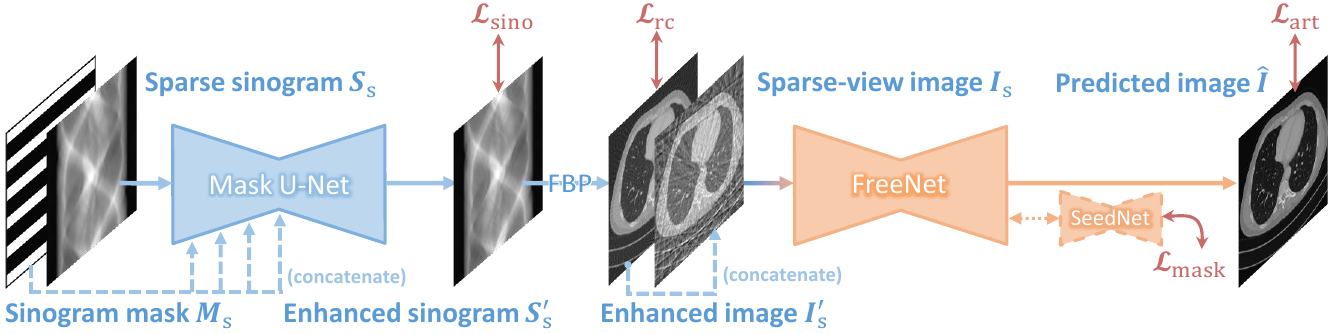}
\caption{Overview of dual-domain counterpart of \freeseed.}
\label{fig:dudofreeseed}
\end{figure}

\subsection{Extending \freeseed to Dual-Domain Framework}
Dual-domain methods are effective in the task of sparse-view CT reconstruction when the sinogram data are available. To further enhance the image reconstruction quality, we extend \freeseed to the dominant dual-domain framework by adding the sinogram-domain sub-network from DuDoNet~\cite{dudonet}, where the resulting dual-domain counterpart shown in Fig.~\ref{fig:dudofreeseed} is called \dudofreeseed. The sinogram-domain sub-network involves a mask U-Net that takes in the linearly interpolated sparse sinogram $\mat{S}_\mathrm{s}$, where a binary sinogram mask $\mat{M}_\mathrm{s}$ that outlines the unseen part of the sparse-view sinogram is concatenated to each stage of the U-Net encoder. The mask U-Net is trained using sinogram loss $\mathcal{L}_\mathrm{sino}$ and radon consistency loss $\mathcal{L}_\mathrm{rc}$. We refer the readers to Lin~\emph{et al.}~\cite{dudonet} for more information.

%% file: secs/experiments.tex
\section{Experiments}
\subsection{Experimental Settings}
We conduct experiments on the dataset of “the 2016 NIH-AAPM Mayo Clinic Low Dose CT Grand Challenge”~\cite{aapm}, which contains 5936 CT slices in 1 mm image thickness from 10 anonymous patients, where a total of 5,410 slices resized to $256\times 256$ resolution are randomly selected for training and the rest 526 slices for testing. Fan-beam CT projection under 120 kVp and 500 mA is simulated using TorchRadon toolbox~\cite{torch-radon}. Specifying the distance from X-ray source to rotation center as 59.5 cm and the number of detectors as 672, we generate sinograms from full-dose images with multiple sparse views $N_\mathrm{v}\in\{18,36,72,144\}$ uniformly sampled from full 720 views covering $[0,2\mathrm{\pi}]$.

The models are implemented in PyTorch~\cite{pytorch} and are trained for 30 epochs with a batch size of 2, using Adam optimizer~\cite{adam} with $(\beta_1,\beta_2)=(0.5,0.999)$ and a learning rate that starts from $10^{-4}$ and is halved every 10 epochs. Experiments are conducted on a single NVIDIA V100 GPU, using the same dataset and setting. 
All sparse-view CT reconstruction methods are evaluated quantitatively in terms of root mean squared error (RMSE), peak signal-to-noise ratio (PSNR) and structural similarity (SSIM)~\cite{quality-assess}.

\begin{figure}[htbp]
\centering
\includegraphics[width=\linewidth]{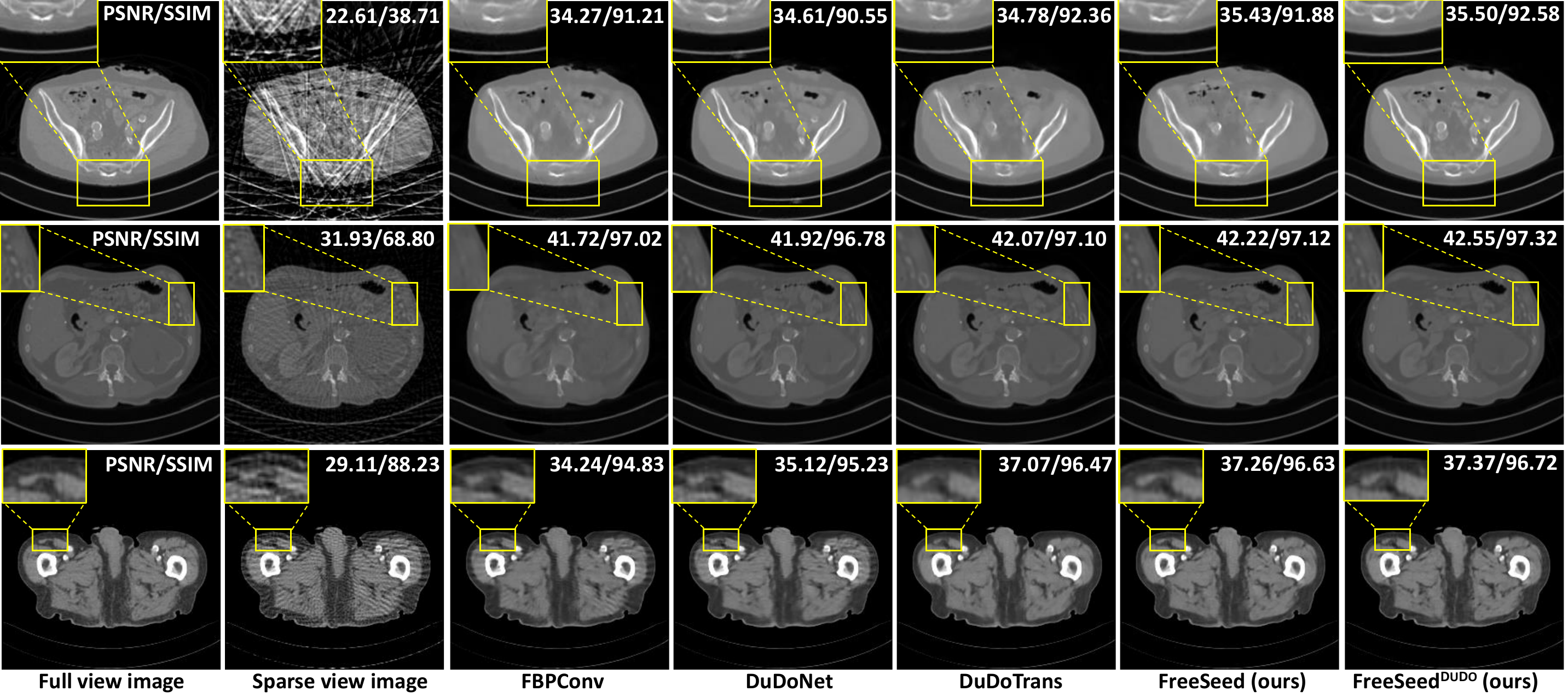}
\caption{Visual comparison of state-of-the-art methods. From top to bottom: $N_\mathrm{v}$=36, 72 and 144; the display windows are [0,2000], [500,3000] and [50,500] HU, respectively.} 
\label{fig:comp-vis}
\end{figure}

\subsection{Overall Performance}

We compare our models (\freeseed and \dudofreeseed) with the following reconstruction methods: direct FBP, \textbf{DDNet}~\cite{ddnet}, \textbf{FBPConv}~\cite{fbpconvnet}, \textbf{DuDoNet}~\cite{dudonet} and \textbf{DuDoTrans}~\cite{dudotrans}. FBPConv and DDNet are image-domain methods, while DuDoNet and DuDoTrans are state-of-the-art dual-domain methods effective for CT image reconstruction. Table~\ref{tab:sota} shows the quantitative evaluation. 

Not surprisingly, we found that the performance of conventional image-domain methods was inferior to the state-of-the-art dual-domain method, mainly due to the failure of removing the global artifacts. We noticed that dual-domain methods underperformed FBPConv when $N_\mathrm{v} = 18$ because of the secondary artifact induced by the inaccurate sinogram restoration in ultra-sparse scenario.

Notably, \freeseed outperformed the dual-domain methods in most settings. Fig.~\ref{fig:comp-vis} provides the visualization results for different methods. In general, \freeseed successfully restored the tiny clinical structures (the spines in the first row, and the ribs in the second row) while achieving more comprehensive artifact removal (see the third row). 
Note that when the sinogram data are available, dual-domain counterpart \dudofreeseed gains further improvements, showing great flexibility of our model.

\subsection{Ablation Study}
Table~\ref{tab:ablation-module} presents the effectiveness of the each component in \freeseed, where seven variants of \freeseed are: 
(1) FBPConv upon which \freenet is built (baseline); 
(2) \freenet without band-pass attention maps nor \seednet guidance $\mathcal{L}_\mathrm{mask}$ (baseline + Fourier); 
(3) FBPConv trained with $\mathcal{L}_\mathrm{mask}$ (baseline + \seednet); 
(4) \freenet trained without $\mathcal{L}_\mathrm{mask}$ (\freenet); 
(5) \freenet trained with simple masked loss $\mathcal{L}_\text{1+mask}=\Vert(\mat{A}_\mathrm{f}-\widehat{\mat{A}})\odot(1+\mat{M})\Vert_2$ ($\text{\freenet}_\text{1+mask}$); 
(6) \freenet trained with $\mathcal{L}_\mathrm{mask}$ using $\ell_1$ norm ($\text{\freeseed}_{\ell_1}$);
(7) \freenet trained with $\mathcal{L}_\mathrm{mask}$ using $\ell_2$ norm, \emph{i.e.}, the full version of our model (\freeseed).

\input{tabs/sotat_aligned2}

\input{tabs/ablation}

By comparing the first two rows of Table~\ref{tab:ablation-module}, we found that simply applying FFC provided limited performance gains. Interestingly, we observed that the advantage of band-pass attention became more pronounced given more views, which can be seen in the last row of Fig.~\ref{fig:amp-artifact} where the attention maps are visualized by averaging all inner radii and bandwidths in different stages of \freenet and calculating the map following Eq.~\eqref{eq:band-pass-gaussian}. Fig.~\ref{fig:amp-artifact} shows that these maps successfully capture the banded pattern of the artifact, especially in the cases of $N_\mathrm{v}=36,72,144$ where artifacts are less entangled with the image content and present a banded shape in the frequency domain. Thus, the band-pass attention maps lead to better convergence. 

The effectiveness of \seednet can be seen by comparing Rows (1) and (3) and also Rows (4) and (7). Both the baseline and \freenet can benefit from the \seednet supervision. Visually, clinical details in the image that are obscured by the heavy artifacts can be further refined by \freenet; please refer to Fig. \ref{fig:supp-refine} in our Supplementary Material for more examples and ablation study. We also found that $\text{\freenet}_\text{1+mask}$ did not provide stable performance gains, probably because directly applying mask on the pixel-wise loss led to discontinuous gradient that brings about sub-optimal results, which, however, can be circumvented with the guidance of \seednet. 
In addition, we trained \freeseed with Eq.~\eqref{eq:loss-model} using $\ell_1$ norm. From the last two rows in Table \ref{tab:ablation-module} we found that $\ell_1$ norm did not ensure stable performance gains when FFC was used.

%% file: tabs/sotat_aligned2.tex
\begin{table}[htbp]
\centering
\caption{Quantitative evaluation for state-of-the-art methods in terms of PSNR [dB], SSIM [\%], and RMSE [$\times10^{-2}$]. 
The best results are highlighted in \textbf{bold} and the second-best results are \underline{underlined}.}
\resizebox{\textwidth}{!}{
\begin{tabular}{l|ccc|ccc|ccc|ccc}
\shline

\multirow{2}{*}{Methods} 
& \multicolumn{3}{c|}{$N_\mathrm{v}=18$} 
& \multicolumn{3}{c|}{$N_\mathrm{v}=36$} 
& \multicolumn{3}{c|}{$N_\mathrm{v}=72$} 
& \multicolumn{3}{c}{$N_\mathrm{v}=144$} \\
& \multicolumn{3}{r|}{PSNR SSIM RMSE}  
& \multicolumn{3}{r|}{PSNR SSIM RMSE}  
& \multicolumn{3}{r|}{PSNR SSIM RMSE}   
& \multicolumn{3}{r}{PSNR SSIM RMSE}    \\
\hline
FBP   
& 22.88 & 36.59 & 7.21 
& 26.44 & 49.12 & 4.78 
& 31.63 & 66.23 & 2.63 
& 38.51 & 86.23 & 1.19 \\

\tabincell{c}{DDNet} 
& 34.07 & 90.63 & 1.99 
& 37.15 & 93.50 & 1.40 
& 40.05 & 95.18 & 1.03 
& 45.09 & 98.37 & 0.56 \\

\tabincell{c}{FBPConv} 
& \textbf{35.04} & 91.19 & \textbf{1.78} 
& 37.63 & 93.65 & 1.32 
& 41.95 & 97.40 & 0.82 
& 45.96 & 98.53 & 0.51 \\

\tabincell{c}{DuDoNet} 
& 34.42 & 91.07 & 1.91 
& 38.18 & 93.45 & 1.24 
& 42.80 & 97.21 & 0.73 
& 47.79 & 98.96 & 0.41 \\

\tabincell{c}{DuDoTrans} 
& 34.89 & 91.08 & 1.81 
& 38.55 & \textbf{94.82} & 1.19 
& 43.13 & 97.67 & 0.70 
& 48.42 & 99.15 & 0.38 \\

\hline\hline
\tabincell{c}{\freeseed} 
& 35.01 & \underline{91.46} & \underline{1.79} 
& \underline{38.63} & 94.46 & \underline{1.18} 
& \underline{43.42} & \underline{97.82} & \underline{0.68} 
& \underline{48.79} & \underline{99.19} & \underline{0.37} \\

\tabincell{c}{\dudofreeseed} 
& \underline{35.03} & \textbf{91.81} & \textbf{1.78} 
& \textbf{38.80} & \underline{94.78} & \textbf{1.16} 
& \textbf{43.78} & \textbf{97.90} & \textbf{0.65} 
& \textbf{49.06} & \textbf{99.23} & \textbf{0.35} \\
\shline
\end{tabular}}
\label{tab:sota}
\end{table}

%% file: tabs/ablation.tex
\begin{table}[htbp]
\centering
\caption{PSNR value of variants of \freeseed. The best results are highlighted in \textbf{bold} and the second-best results are \underline{underlined}.}
\begin{tabular*}{0.8\linewidth}{@{\extracolsep{\fill}}l|cccc}
\shline
\multicolumn{1}{c|}{Variants} & \multicolumn{1}{c}{$N_\mathrm{v}$=18} & \multicolumn{1}{c}{$N_\mathrm{v}$=36} & \multicolumn{1}{c}{$N_\mathrm{v}$=72} & \multicolumn{1}{c}{$N_\mathrm{v}$=144}\\
\hline
(1) baseline       
& \textbf{35.04}
& 37.63 
& 41.95 
& 45.96
\\
(2) baseline + Fourier  
& 34.78 
& 38.23 
& 42.33 
& 47.32 
\\
(3) baseline + \seednet  
& 34.49 
& 38.35 
& 42.89 
& 48.64 
\\
(4) \freenet
& 34.77 
& 38.42 
& 43.06 
& 48.63 
\\
(5) $\text{\freenet}_\text{1+mask}$
& 34.54 
& 38.17 
& 42.94 
& 48.73 
\\
(6) $\text{\freeseed}_{\ell_1}$
& 34.79 
& 38.45 
& 43.06 
& \textbf{49.00} 
\\
(7) \freeseed (ours)
& \underline{35.01} 
& \textbf{38.63} 
& \textbf{43.42} 
& \underline{48.79} 
\\
\shline
\end{tabular*}
\label{tab:ablation-module}
\end{table}

%% file: secs/conclusion.tex
\section{Conclusion}
In this paper, we propose \freeseed, a simple yet effective image-domain method for sparse-view CT reconstruction. \freeseed incorporates Fourier knowledge into the reconstruction network with learnable band-pass attention for a better grasp of the globally dispersed artifacts, and is trained using a self-guided artifact refinement network to further refine the heavily damaged image details.
Extensive experiments show that both \freeseed and its dual-domain counterpart outperformed the state-of-the-art methods. In future, we will explore FFC-based network for sinogram interpolation in sparse-view CT reconstruction.

%% file: secs/supp.tex
\begin{algorithm}[htbp]
\caption{The training process of \freeseed.}
\label{alg:training}
\begin{algorithmic}[1]
    \REQUIRE FreeNet: $\Theta$;  SeedNet: $\Phi$; Sparse-view CT images: $\mat{I}_\mathrm{s}$; Full-view CT images: $\mat{I}_\mathrm{f}$; Maximum number of iterations: $n$.
    \FOR{$i$ \textbf{in} range($n$)}
    \STATE $\widehat{\mat{I}}^{(i)},\widehat{\mat{A}}^{(i)}=\Theta^{(i)}(\mat{I}_\mathrm{s}),$ \\
    \STATE $\widetilde{\mat{I}}^{(i)}=\Phi^{(i)}(\widehat{\mat{I}}^{(i)}),$ \\
    \STATE Update $\Phi^{(i+1)}$ using $\mathcal{L}_\mathrm{mask}$ with $\mat{I}_\mathrm{f}, \widetilde{\mat{I}}^{(i)}, \widehat{\mat{A}}^{(i)}$.
    \hfill \# SeedNet training
    \STATE $\widetilde{\mat{I}}^{(i+1)}=\Phi^{(i+1)}(\widehat{\mat{I}}^{(i)}),$ \\
    \STATE Update $\Theta^{(i+1)}$ using $\mathcal{L}_\mathrm{total}$ with $\mat{I}_\mathrm{f}, \widetilde{\mat{I}}^{(i+1)}, \widehat{\mat{A}}^{(i)}, \widehat{\mat{I}}^{(i)}$.
    \hfill \# FreeNet training
    \ENDFOR
    \RETURN $\Theta^{(n)}$
\end{algorithmic}
\end{algorithm}

\begin{figure}[htbp]
\centering
\subfigure[PSNR]
{\includegraphics[width=0.32\linewidth]{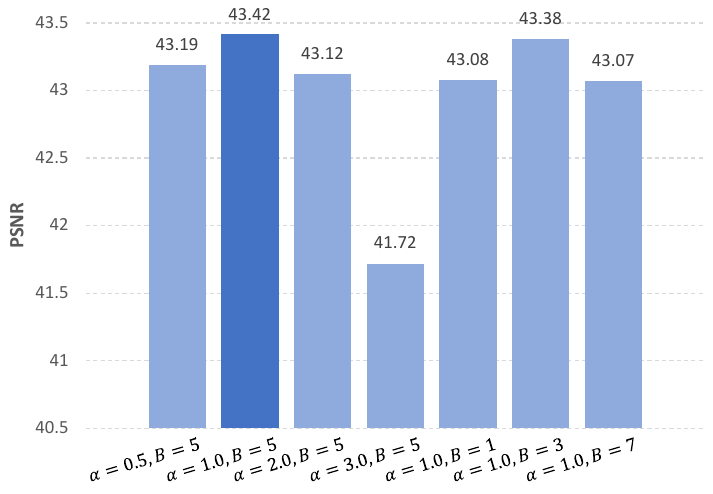}}
\label{fig:supp-ab-psnr}
\subfigure[SSIM]
{\includegraphics[width=0.32\linewidth]{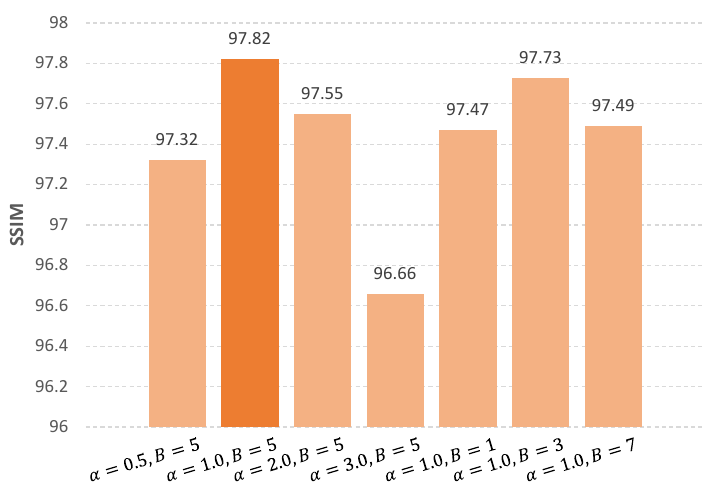}}
\label{fig:supp-ab-ssim}
\subfigure[RMSE]
{\includegraphics[width=0.32\linewidth]{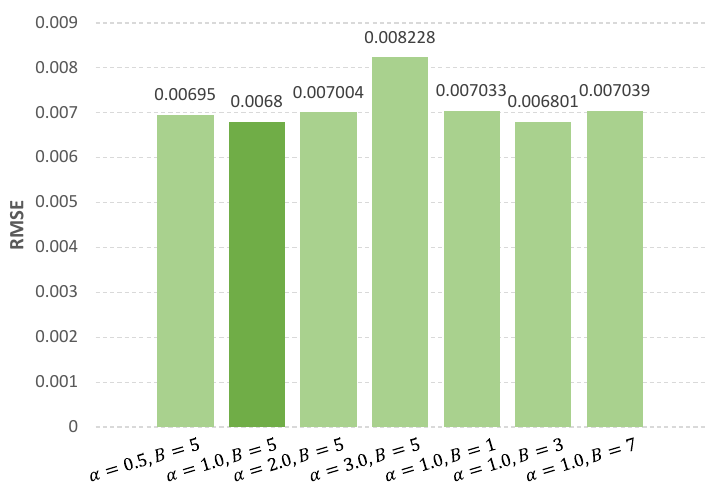}}
\label{fig:supp-ab-rmse}
\caption{Quantitative evaluation of variants under $N_\mathrm{v}=72$ using different coefficients $\alpha$ and number of Fourier convolution blocks $B$ in the SeedNet, with the full version of the proposed model denoted by “$\alpha=1.0,B=5$”.}
\label{fig:supp-ab}
\end{figure}

\begin{figure}[htbp]
\centering
\includegraphics[width=\linewidth]{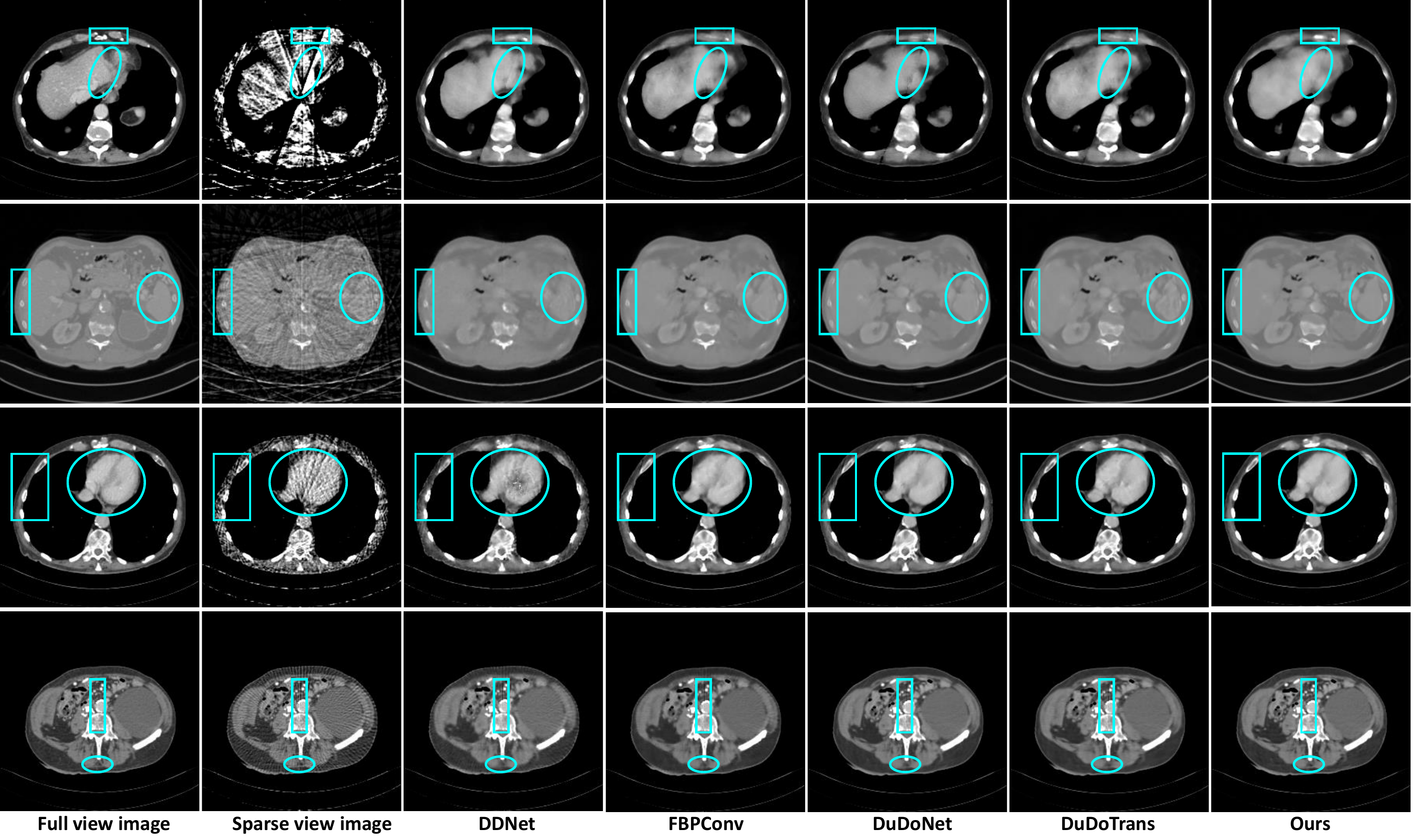}
\caption{More visualization samples of state-of-the-art methods. From top to bottom: $N_\mathrm{v}$=18, 36, 72 and 144; the display window for $N_\mathrm{v}=36$ is [0,2000] HU while others are [50,500] HU.} 
\label{fig:supp-vis}
\end{figure}

\begin{figure}[htbp]
\centering
\includegraphics[width=\linewidth]{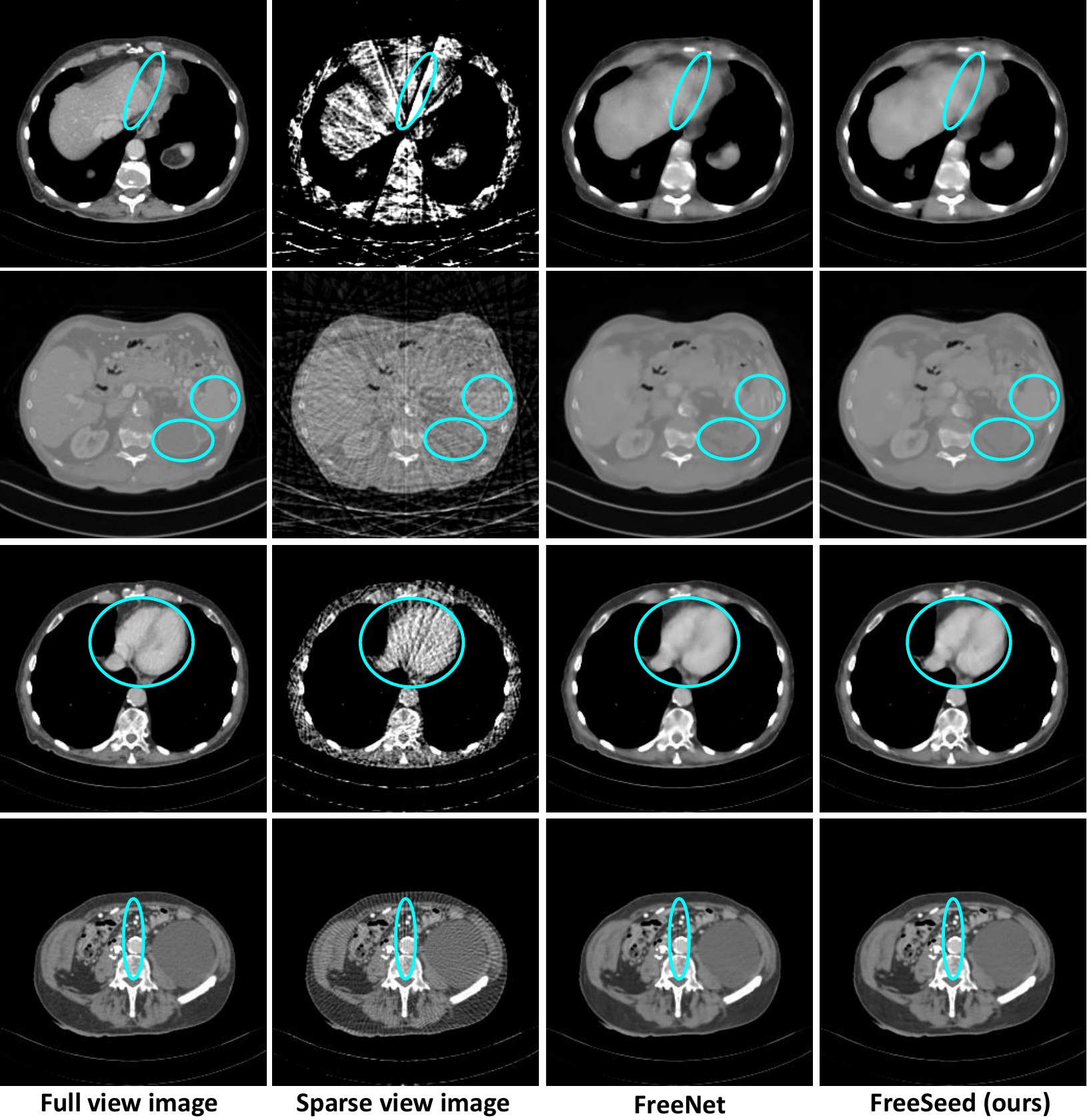}
\caption{Visualization samples showing the effects of SeedNet, where the last column (the full version of the proposed model) recovers more detail. From top to bottom: $N_\mathrm{v}$=18, 36, 72 and 144; the display window for $N_\mathrm{v}=36$ is [0,2000] HU while others are [50,500] HU.} 
\label{fig:supp-refine}
\end{figure}

%% file: main.bbl
\begin{thebibliography}{10}
\providecommand{\url}[1]{\texttt{#1}}
\providecommand{\urlprefix}{URL }
\providecommand{\doi}[1]{https://doi.org/#1}

\bibitem{ffc}
Chi, L., Jiang, B., Mu, Y.: Fast {Fourier} convolution. In: Advances in Neural
  Information Processing Systems. vol.~33, pp. 4479--4488 (2020)

\bibitem{residual-manifold}
Han, Y.S., Yoo, J., Ye, J.C.: Deep residual learning for compressed sensing ct
  reconstruction via persistent homology analysis. arXiv preprint
  arXiv:1611.06391  (2016)

\bibitem{resnet}
He, K., Zhang, X., Ren, S., Sun, J.: Deep residual learning for image
  recognition. In: Proceedings of the IEEE/CVF Conference on Computer Vision
  and Pattern Recognition. pp. 770--778 (2016)

\bibitem{fbpconvnet}
Jin, K.H., McCann, M.T., Froustey, E., Unser, M.: Deep convolutional neural
  network for inverse problems in imaging. IEEE Transactions on Image
  Processing  \textbf{26}(9),  4509--4522 (2017)

\bibitem{adam}
Kingma, D.P., Ba, J.: {Adam}: A method for stochastic optimization. arXiv
  preprint arXiv:1412.6980  (2014)

\bibitem{sinogram-interp}
Lee, H., Lee, J., Kim, H., Cho, B., Cho, S.: Deep-neural-network-based sinogram
  synthesis for sparse-view {CT} image reconstruction. IEEE Transactions on
  Radiation and Plasma Medical Sciences  \textbf{3}(2),  109--119 (2018)

\bibitem{dudonet}
Lin, W.A., Liao, Haofu abd~Peng, C., Sun, X., Zhang, J., Luo, J., Chellappa,
  R., Kevin, Z.S.: {DuDoNet}: Dual domain network for {CT} metal artifact
  reduction. In: Proceedings of the IEEE/CVF Conference on Computer Vision and
  Pattern Recognition. pp. 10512--10521 (2019)

\bibitem{aapm}
McCollough, C.: {TU-FG-207A-04}: Overview of the low dose {CT} grand challenge.
  Medical Physics  \textbf{43}(6),  3759--3760 (2016)

\bibitem{alara}
Miller, D.L., Schauer, D.: The alara principle in medical imaging. Philosophy
  \textbf{44},  595--600 (1983)

\bibitem{pytorch}
Paszke, A., Gross, S., Massa, F., Lerer, A., Bradbury, J., Chanan, G., Killeen,
  T., Lin, Z., Gimelshein, N., Antiga, L., et~al.: Pytorch: An imperative
  style, high-performance deep learning library. Advances in Neural Information
  Processing Systems  \textbf{32} (2019)

\bibitem{torch-radon}
Ronchetti, M.: {TorchRadon}: Fast differentiable routines for computed
  tomography. arXiv preprint arXiv:2009.14788  (2020)

\bibitem{lama}
Suvorov, R., Logacheva, E., Mashikhin, A., Remizova, A., Ashukha, A.,
  Silvestrov, A., Kong, N., Goka, H., Park, K., Lempitsky, V.:
  Resolution-robust large mask inpainting with {Fourier} convolutions. In: 2022
  IEEE/CVF Winter Conference on Applications of Computer Vision. pp. 2149--2159
  (2022)

\bibitem{dudotrans}
Wang, C., Shang, K., Zhang, H., Li, Q., Zhou, S.K.: {DuDoTrans}: Dual-domain
  transformer for sparse-view {CT} reconstruction. In: Machine Learning for
  Medical Image Reconstruction. pp. 84--94 (2022)

\bibitem{ct-outlook}
Wang, G., Yu, H., De~Man, B.: An outlook on {X}-ray {CT} research and
  development. Medical Physics  \textbf{35}(3),  1051--1064 (2008)

\bibitem{quality-assess}
Wang, Z., Bovik, A.C., Sheikh, H.R., Simoncelli, E.P.: Image quality
  assessment: from error visibility to structural similarity. IEEE Transactions
  on Image Processing  \textbf{13}(4),  600--612 (2004)

\bibitem{drone}
Wu, W., Hu, D., Niu, C., Yu, H., Vardhanabhuti, V., Wang, G.: {DRONE}:
  Dual-domain residual-based optimization network for sparse-view {CT}
  reconstruction. IEEE Transactions on Medical Imaging  \textbf{40}(11),
  3002--3014 (2021)

\bibitem{swinfir}
Zhang, D., Huang, F., Liu, S., Wang, X., Jin, Z.: {SwinFIR}: Revisiting the
  {SwinIR} with fast {Fourier} convolution and improved training for image
  super-resolution. arXiv preprint arXiv:2208.11247  (2022)

\bibitem{ddnet}
Zhang, Z., Liang, X., Dong, X., Xie, Y., Cao, G.: A sparse-view {CT}
  reconstruction method based on combination of {DenseNet} and deconvolution.
  IEEE Transactions on Medical Imaging  \textbf{37}(6),  1407--1417 (2018)

\end{thebibliography}
